\documentclass[12pt]{iopart}
\usepackage{iopams,color,graphicx,multirow,acronym}
\usepackage[colorlinks,linkcolor=blue,citecolor=blue,urlcolor=blue ]{hyperref}
\newcommand{\be}{\begin{equation}}
\newcommand{\ee}{\end{equation}}
\newcommand{\bea}{\begin{eqnarray}}
\newcommand{\eea}{\end{eqnarray}}
\newcommand{\um}{{\rm m}}
\newcommand{\ukm}{{\rm km}}
\newcommand{\ucm}{{\rm cm}}
\newcommand{\umm}{{\rm mm}}

\newcommand{\upm}{{\rm pm}}

\newcommand{\uHz}{{\rm Hz}}

\newcommand{\umK}{{\rm mK}}

\newcommand{\us}{{\rm s}}

\newcommand{\muN}{\mu{\rm N}}

\def\ccbx#1{{#1}}
\def\ccrx#1{{#1}}
\newcommand{\Msun}{{\rm M}_\odot}

\acrodef{CMB}{cosmic microwave background}
\acrodef{EMRI}{extreme mass ratio inspiral}
\acrodef{FIM}{Fisher information matrix}
\acrodef{GR}{general relativity}
\acrodef{GW}{gravitational wave}
\acrodef{LISA}{Laser Interferometer Space Antenna}
\acrodef{LIGO}{Laser Interferometer Gravitational-Wave Observatory}
\acrodef{PTA}{pulsar timing array}
\acrodef{SNR}{signal-to-noise ratio}
\acrodef{GCBs}{Galactic ultra-compact binaries}
\acrodef{SBHBs}{stellar-mass black hole binaries}
\acrodef{MBHBs}{massive black hole binaries}
\acrodef{MBHB}{massive black hole binary}
\acrodef{DWD}{double white dwarf}
\acrodef{SGWB}{stochastic gravitational wave background}
\acrodef{VBs}{verification binaries}
\newcommand{\TRC}{TianQin Research Center for Gravitational Physics \& School of Physics and Astronomy, Sun Yat-sen University (Zhuhai Campus), Zhuhai 519082, P.R. China}

\newcommand{\CGE}{Centre for Gravitational Experiments, School of Physics, MOE Key Laboratory of Fundamental Physical Quantities Measurement \& Hubei Key Laboratory of Gravitation and Quantum Physics, PGMF, Huazhong University of Science and Technology, Wuhan 430074, P. R. China}

\newcommand{\DFH}{DFH Satellite Co., Ltd., Beijing 100094, P.R. China}
\newcommand{\ICE}{Beijing Institute of Control Engineering, Beijing 100094, P.R. China}
\newcommand{\ISSE}{Beijing Institute of Spacecraft System Engineering, Beijing 100094, P.R. China}
\newcommand{\YNO}{Yunnan Observatories, Chinese Academy of Sciences, Kunming 650011, China}
\newcommand{\SISSA}{SISSA, Via Bonomea 265, 34136 Trieste, Italy and INFN Sezione di Trieste \& IFPU - Institute for Fundamental Physics of the Universe, Via Beirut 2, 34014 Trieste, Italy}
\newcommand{\USMB}{Dipartimento di Fisica ``G. Occhialini", Universit\'{a} degli Studi Milano Bicocca, Piazza della Scienza 3, I-20126 Milano, Italy}
\newcommand{\UoB}{School of Physics and Astronomy, University of Birmingham, Birmingham B15 2TT, United Kingdom}
\newcommand{\PKU}{Center for High Energy Physics \& Department of Physics and State Key Laboratory of Nuclear Physics and Technology, Peking University, No.5 Yiheyuan Rd, Beijing 100871, P.R. China}
\newcommand{\PMO}{Purple Mountain Observatory, Chinese Academy of Sciences, Nanjing 210023 \& School of Astronomy and Space Science, University of Science and Technology of China, Hefei, Anhui 230026, P.R. China}
\newcommand{\NAOC}{National Astronomical Observatories, Chinese Academy of Sciences, Beijing 100012, China}
\newcommand{\UCAS}{School of Astronomy and Space Science, University of Chinese Academy of Sciences, Beijing 100049, China}
\newcommand{\AEI}{Max-Planck-Institut f{\"u}r Gravitationsphysik (Albert-Einstein-Institut), D-30167 Hannover, Germany}
\newcommand{\LBU}{Leibniz Universit{\"a}t Hannover, D-30167 Hannover, Germany}
\newcommand{\SAI}{Lomonosov Moscow State University, Sternberg Astronomical Institute, Moscow 119992, Russia}

\newcommand{\Glasgow}{SUPA, School of Physics and Astronomy, University of Glasgow, Glasgow G12 8QQ, UK}
\newcommand{\PdT}{Former Faculty, Politecnico di Torino, Corso Duca degli Abruzzi 24, 10129, Torino, Italy}
\newcommand{\SJTU}{School of Aeronautics and Astronautics, Shanghai Jiao Tong University, Shanghai 200240, China}
\newcommand{\SWU}{School of Physical Science and Technology, Southwest University, Chongqing 400715, China}
\newcommand{\NEU}{Department of Physics, College of Sciences \& MOE Key Laboratory of Data Analytics and Optimization for Smart Industry, Northeastern University, Shenyang 110819, China}

\begin{document}

\title{\ccrx{The TianQin project: current progress on science and technology}}

\author{Jianwei Mei$^1$,
Yan-Zheng Bai$^2$,
Jiahui Bao$^1$,
Enrico Barausse$^{7}$,
Lin Cai$^2$,
Enrico Canuto$^{8}$,
Bin Cao$^1$,
Wei-Ming Chen$^1$,
Yu Chen$^1$,
Yan-Wei Ding$^1$
% ...
}
\address{$^1$\TRC}
\address{$^2$\CGE}
% ...
\address{$^7$\SISSA}
\address{$^{8}$\PdT}
% ...
\ead{$^a$junluo@mail.sysu.edu.cn}
\vspace{10pt}

\date{\today}

\begin{abstract}
TianQin is a \ccbx{planned} space-based \ac{GW} observatory \ccbx{consisting} of three earth orbiting satellites with \ccbx{an} orbital radius of about $10^5~\ukm$.
The satellites \ccrx{will} form a equilateral triangle constellation \ccbx{the plane of which} is nearly perpendicular to the ecliptic plane.
TianQin aims to detect \acp{GW} between $10^{-4}~\uHz$ and $1~\uHz$ that can be generated by a wide variety of important astrophysical and cosmological sources, including the inspiral of Galactic ultra-compact binaries, the inspiral of stellar-mass black hole binaries, \ccbx{extreme mass ratio inspirals}, the merger of massive black hole binaries, and possibly the energetic processes in the very early universe or exotic sources such as cosmic strings.
In order to start \ccbx{science operations} around 2035, a roadmap called the 0123 plan is being used to bring the key technologies of TianQin to maturity, supported \ccbx{by} the construction of a series of research facilities on the ground.
Two major projects of the 0123 plan are being \ccbx{carried out.
In this process,} the team has created a new generation $17~\ucm$ single-body hollow corner-cube retro-reflector which has been launched with the QueQiao satellite on 21 May 2018\ccbx{;}
a new laser ranging station equipped with a $1.2~\um$ telescope has been constructed and the station has successfully ranged to all the five retro-reflectors on the Moon\ccbx{;}
and the TianQin-1 experimental satellite has been launched on 20 December 2019 and the first round result shows that the satellite has exceeded all of its mission requirements.
\end{abstract}



\section{Introduction}


Major activities in \ac{GW} detection are expected around 2035\ccbx{. By then} the network of ground-based \ac{GW} detectors should have detected thousands of \ac{GW} events and the pulsar timing array and the cosmic microwave background \ccbx{experiments} could have made \ccbx{historic} breakthroughs.
Brand new space-based \ac{GW} observatories, such as LISA \cite{Audley:2017drz} and DECIGO \cite{Sato:2017dkf,Isoyama:2018rjb}, could also join the effort.
Proposed in 2014, the TianQin project aims to launch the space-based \ac{GW} observatory, TianQin, around 2035 and to detect \acp{GW} between $10^{-4}~\uHz$ and $1~\uHz$ \cite{Luo:2015ght}.
TianQin is unique in several \ccbx{respects}: it is the \ccbx{only planned detector that uses} geocentric orbits \cite{Hellings:1996vz,2012cosp...39.1890S,Tinto:2014eua}, \ccbx{like the others} it uses three satellites to form a \ccbx{equilateral} triangular constellation\ccbx{,} but the constellation plane is nearly perpendicular to the ecliptic plane, and its frequency sensitivity band is in between those of LISA and DECIGO, overlapping with that of LISA near $10^{-4}~\uHz$ and with that of DECIGO near $1~\uHz$.

As first published in \cite{Luo:2015ght}, the concept of TianQin envisions a \ccbx{equilateral} triangle constellation of three drag-free satellites orbiting the Earth with an orbital radius of about $10^5$ km \cite{Ye:2019txh,Hu:2018yqb,doi:10.1142/S021827182050056X}, with the detector orientation, i.e. the normal vector to the constellation plane of the detector, pointing toward RX J0806.3+1527 (also known as HM Cancri or HM Cnc, hereafter J0806 \cite{Strohmayer:2005uc}).
The satellites will be carefully controlled to provide \ccbx{an ultra-clean} and stable environment for \ccbx{its} scientific operation, allowing gravity to take full governance of the motion of a set of test masses residing in the satellites and allowing laser interferometry to reach extremely high precision. In this way, the variation in the distance between the test masses (partially caused by \acp{GW}) can be measured with the inter-satellite laser interferometry.

Some basic parameters of TianQin \ccbx{are} listed in Table \ref{tab:configuration}.

\renewcommand\arraystretch{1.5}   
\begin{table*}[b]
\caption{Basic parameters of TianQin.} \label{tab:configuration}
\begin{tabular}{l|l}
\hline
Parameters              & Value  \\
\hline
Type of orbit           & Geocentric \\
Number of satellites    & N=3 \\
Arm length              & $L = \sqrt{3}\times10^5~\ukm$ \\
Displacement measurement noise         & $S_x^{1/2} = 1\times10^{-12}~\um/\uHz^{1/2}$ \\
Residual acceleration noise      & $S_a^{1/2} = 1\times10^{-15}~\um~\us^{-2}/\uHz^{1/2}$ \\
\hline
\end{tabular}\\
\end{table*}

\section{Scientific \ccbx{prospects}}\label{sec:science}

A systematic effort has been \ccbx{undertaken} to assess the expected science output of TianQin \cite{Wang:2019ryf,Feng:2019wgq,Shi:2019hqa,Bao:2019kgt,Liu:2020eko,Huang:2020rjf,Fan:2020zhy,Liang:2020sto,Zhu:2020}. In these works, the following sensitivity curve has been used \cite{Luo:2015ght,Hu:2018yqb,Lu:2019log},
\be S_n(f) = \frac{10}{3L^2}\left[S_x+\frac{4S_a}{(2\pi f)^4}\left(1+\frac{10^{-4}{\rm Hz}}{f}\right) \right] \times\Big[1+0.6\Big(\frac{f}{f_\ast}\Big)^2\Big]\,,
\label{eq:S_n^SA} \ee
where $S_x$ and $S_a$ are given in Table \ref{tab:configuration}, and $f_\ast=c/(2\pi L)\approx0.28~\uHz$ is the transfer frequency.
Note (\ref{eq:S_n^SA}) \ccbx{describes} a ballpark goal and the sensitivity curve is expected to be refined over time. \ccrx{We assume 5 years of mission lifetime for TianQin. With its baseline concept \cite{Luo:2015ght}, TianQin is expected to operate in a consecutive ``three-month on + three-month off" mode, which means that TianQin will firstly observe continuously for three months, and then be put into a safe mode for the next three months before starting observation again, in order to cope with the variation of thermal load on the satellites. In this scheme, the total duration of data acquisition will be of 2.5 years for a mission lifetime of 5 years. Although various possibilities are being explored to increase the fraction of total observation time, we use the baseline concept of TianQin to get a (presumably conservative) assessment of the expected science output.}

The major sources \ccbx{expected for} TianQin include the inspiral of \ac{GCBs}, the inspiral of \ac{SBHBs}, \ccbx{\acp{EMRI}}, the merger of \ac{MBHBs}, and possibly also energetic processes in the very early universe or exotic sources such as cosmic strings \cite{Hu:2017yoc}.
With the detection of \acp{GW} from such sources, TianQin is expected to provide key information on the astrophysical history of galaxies and black holes, the dynamics of dense star clusters and galactic centers, the nature of gravity and black holes, the expansion \ccbx{history} of the universe, and possibly also the fundamental physics related to the energetic processes \ccbx{in} the early universe.

\begin{figure}
\centerline{\includegraphics[width=0.6\linewidth]{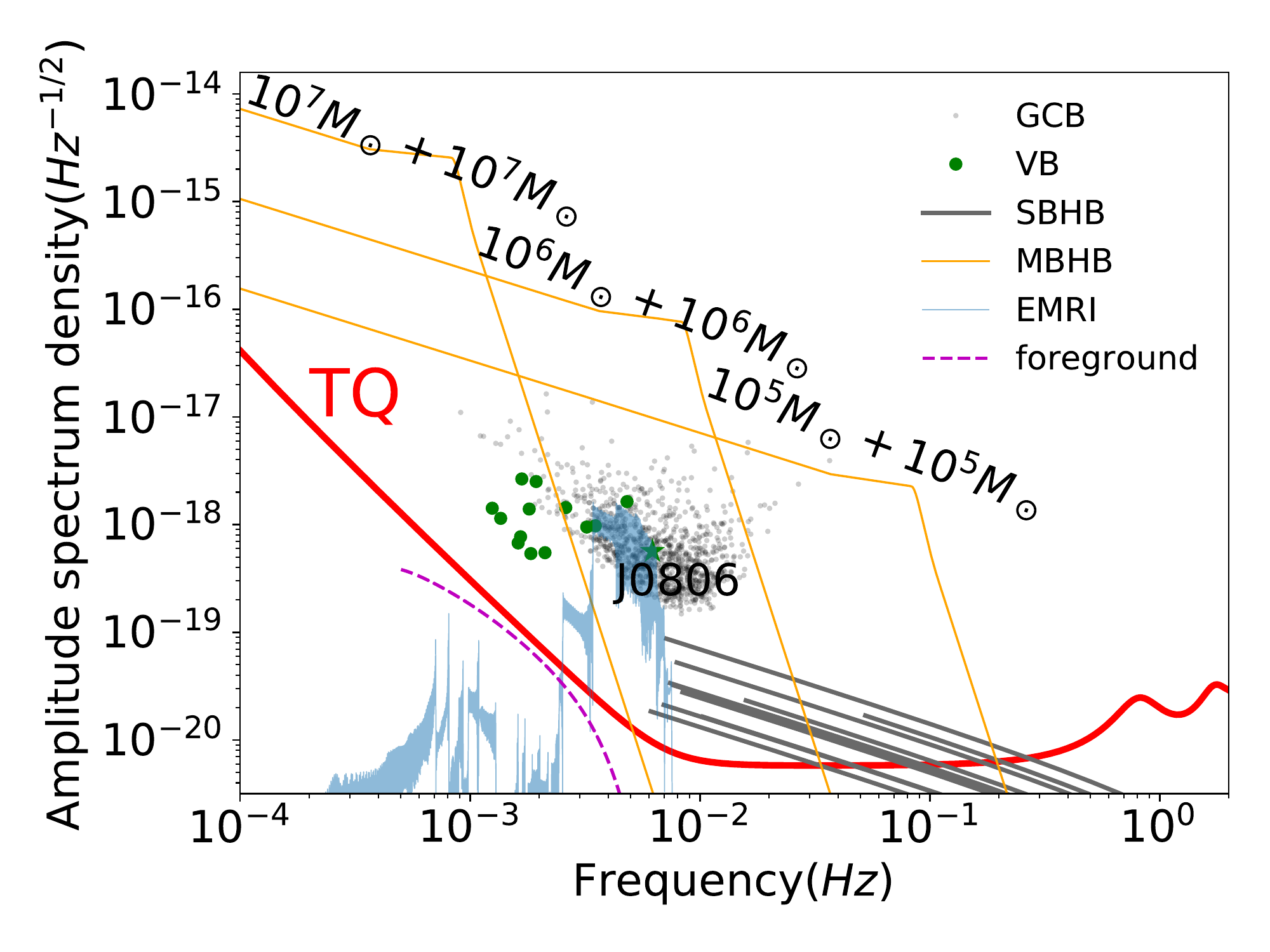}}
\caption{\ccrx{Expected astrophysical sources} for TianQin. 
In this plot, the information of the detector response is included in the signal strain of each source and the instrumental noise (the solid red curve) is approximated by dividing $S_n(f)$ \ccbx{by} $10/3$, which accounts for the geometric configuration of TianQin, as well as the location- and
polarisation-sensitive response in the low frequency limit.}
\label{fig:source}
\end{figure}

A summary of the \ccrx{expected astrophysical sources} that can be detected with TianQin is illustrated in Figure \ref{fig:source}.

\subsection{\ccbx{Galactic ultra-compact binaries (GCBs)}}

\ac{GCBs} are likely the most numerous sources for \ccrx{a space-based GW detector such as} TianQin \cite{Evans:1987qa,1987A&A...176L...1L,Hils:1990vc,Nelemans:2013yg,Huang:2020rjf}.
The detection of \ac{GCBs} is of great importance for astrophysics and fundamental physics.
For astrophysics, \ccbx{\ac{GW} observations are} \ccrx{likely to detect more} \ac{GCBs} with \ccbx{ultra-short periods ($< 1$\,hour)} than the electromagnetic observatories \cite{Rebassa-Mansergas:2018mtp}, \ccbx{and} thus can not only \ccbx{subject} the formation theories of \ac{GCBs} to precise test but can also help \ccbx{to} study the matter distribution and the structure of the Galaxy \ccrx{\cite{Adams:2012qw,Korol:2018wep,Wilhelm:2020qjc}}.
The \acp{GW} from \ac{GCBs} can also be used to study fundamental physics, such as \ccbx{helping to constrain} extra radiation channels or extra polarisation modes of \acp{GW} \cite{Littenberg:2018xxx,Philippoz:2017ywb}. \ccrx{It has been shown that TianQin can facilitate studies on the formation of neutron star systems through accretion-induced collapse in white-dwarf binaries \cite{Wang:2020pzc}. TianQin also has the potential to detect \acp{GW} from deformed compact stars \cite{Rodriguez:2019ucq}.}

Electromagnetic observation has identified dozens of \ac{GCBs} that have orbital periods shorter than a few hours \cite{Strohmayer:2005uc,Brown:2011gq,kil14,kup18,ram18,bur19}.
These \ac{GCBs}, once detected, can become very good calibration sources for the detector and they are referred to as \ccbx{\ac{VBs}}.
For example,  \ccrx{J0806} has the shortest known period \ccbx{of} all \ac{GCBs} \cite{Strohmayer:2005uc}.
TianQin can detect J0806 with a \ac{SNR} \ccbx{of} 5 \ccbx{already} after two days of observation.
TianQin can also detect 12 \ac{VBs}, for some of which the \ac{GW} amplitude can be determined up to the 5\% level if the amplitude and the inclination angle are not precisely known \ccbx{while other parameters are assumed to be constrained by electromagnetic studies}.

By using a synthetic population of \ac{GCBs}, we find that TianQin can resolve about $8.7\times10^3$ \ac{GCBs}, and for these events, the uncertainties in the parameter estimation center around: $\Delta  P/P = 31.6 \times 10^{-8}$, $\Delta \mathcal{A}/\mathcal{A} = 0.10$, $\Delta \cos\iota = 0.05$, and $\Delta \Omega_S = 7.94$ deg$^2$, where $P$ and ${\cal A}$ are the period and amplitude of the \ac{GW} signal, and $\iota$ and $\Omega_S$ are the inclination angle and the sky localization (the solid angle in which the source is located) of the source, respectively.
Among the \ac{GCBs} resolvable with TianQin, about 38\% can be localized to better than 1 deg$^2$.
In the lower frequency range, the incoherent addition of \ccbx{a} vast number of \ac{GCBs} will form a confusion noise, or sometimes referred to as \ccbx{a} foreground.
An order of magnitude estimation shows that TianQin may also detect one double white dwarf merger event.
More detail on the detection of \ac{GCBs} with TianQin can be found in \cite{Huang:2020rjf}.

\subsection{\ccbx{Stellar-mass black hole binaries (SBHBs)}}

Another type of quite promising \ccbx{source} for TianQin \ccbx{is} \ac{SBHBs}. The discoveries made by the LIGO Scientific Collaboration and Virgo Collaboration have revealed that there exist many \ac{SBHBs} \cite{LIGOScientific:2018mvr}. The merger of \ac{SBHBs} will generate \acp{GW} with frequencies \ccbx{of order 100Hz. However, years to months before the final merger,} the \acp{GW} are in the milli-Hertz regime, making \ccbx{\ac{SBHBs}} interesting sources for TianQin. \ccbx{\ac{SBHBs}} are ideal objects for multi-band \ac{GW} detection, which can not only help better detect and measure the sources themselves but also make the \ac{SBHBs} powerful laboratories for cosmology and fundamental physics, \ccbx{as illustrated by their potential use as} standard sirens for cosmology \cite{Colpi:2016fup} and to greatly improve the constraints on certain parameters of modified gravity \ccbx{theories} \cite{Sesana:2016ljz}.
\ccrx{It has been pointed out that TianQin can play an important role in detecting sources in the deci-Hertz frequency range \cite{Chen:2017gfm,Samsing:2019dtb}, and that it is possible to learn about the eccentricity distribution and formation channels by counting the number of \ac{SBHBs} observed with TianQin \cite{Randall:2019znp}. Some work on the potential of using TianQin to test \ac{GR} with \ac{GW} signals from \ac{SBHBs} can be found in \cite{Carson:2019yxq}.}

There is large uncertainty in the mass distribution of the \ac{SBHBs}. As many as five models have been adopted and calibrated by the LIGO Scientific Collaboration and Virgo Collaboration to study the population properties of SBHBs \cite{LIGOScientific:2018jsj}.
The expected \ccbx{number of detections} depends strongly on the \ac{SNR} threshold.
Searching using \ccbx{template banks} requires that the \ac{SNR} threshold for TianQin is about 12.
With this threshold, there is \ccbx{the} possibility that \ccbx{a} single digit number of \ac{SBHBs} events can be detected with TianQin.
With the advancement of data analysis techniques, the requirement on the \ac{SNR} threshold may become less stringent by the time when TianQin flies.
If the \ac{SNR} threshold is lowered to 8, then a single digit \ccbx{number of detections becomes} very probable and there is \ccbx{the possibility that the number of detections} can reach a few dozens. For these events, the probability distribution in redshift peaks at $z\sim0.05$, and the probability distribution in the logarithm of the total mass peaks at \ccbx{$M\sim70~\Msun$}.
\ccbx{Concerning} the precision of parameter estimation, the probability distributions of the (relative) uncertainties in the coalescing time, the sky localization, the chirp mass, the eccentricity and the luminosity distance peak at $\Delta t_c\sim0.1~\us$, $\Delta\bar\Omega_S\sim0.1$ deg$^2$, $\Delta{\cal M}/{\cal M}\sim10^{-7}$, $\Delta e_0/e_0\sim10^{-4}$ and $\Delta D_L/D_L\sim20\%$, respectively. For a typical merger, the error volume is small enough to contain only the host galaxy, while it  can be as small as 2 Mpc$^3$ in the most optimal cases. More detail on the detection of \ac{SBHBs} with TianQin can be found in \cite{Liu:2020eko}.

\subsection{\ccbx{Massive black hole binaries (MBHBs)}}

Most galaxies in the universe host a massive black hole at their centers \cite{1971MNRAS.152..461L,1982MNRAS.200..115S,Kormendy:1995er,Gultekin:2009qn}. The masses of the central black holes are intimately related to the intrinsic parameters of the corresponding host galaxies. When galaxies collide, some of the massive black holes form gravitationally bound pairs, i.e., \ccbx{\ac{MBHBs}}, and eventually merge due to the loss of energy and angular momentum through \ac{GW} radiation.
The \ccbx{mergers} of massive black holes are extremely strong sources of \acp{GW}, which travel through the universe for billions of years but can still preserve significant strength. Some of the signals will be detected by TianQin.
The detection of \acp{GW} from the merger of \ac{MBHBs} can help reveal the origin and the formation channels of \ccbx{massive black holes}, provide a new method to study the expansion of the universe at high redshift and test various aspects of \ac{GR} or the nature of \ccbx{black holes} in the strong field regime \cite{Bao:2019kgt,Berti:2018cxi,Shi:2019hqa}.
\ccrx{With \ac{GW} signals from \ac{MBHBs}, there is \ccbx{the} possibility to use TianQin to explore non-singular black holes \cite{Liu:2020ddo} and to constrain the mass spectrum and number of axionlike fields by measuring black hole spins \cite{Stott:2018opm}.}

How the massive black holes are formed \ccbx{throughout} the history of the universe is still under debate.
For example, it is unclear whether \ccbx{massive black holes} are formed from the direct collapse of a massive cloud (the heavy seed model) or the remnant of population III stars (the light seed model) \cite{Woods:2018lty}, \ccrx{and various models have been proposed to depict the evolution of massive black holes \cite{Barausse:2012fy,Klein:2015hvg,Barausse:2020mdt,Jiang_2019}.}
By using five different models for the event rate of \ac{MBHBs}, it has been found the expected detection rate varies significantly from one model to another, with the most pessimistic model predicting less than $0.1$ detection per year, \ccbx{and} the most optimistic predicting nearly $60$ detections per year, while other intermediate models \ccbx{predict} ${\cal O}(1\sim10)$ detections per year.
If \ac{MBHBs} with nearly equal component masses of the order $10^4\sim10^5~\Msun$ at redshift 15 are detected with TianQin, then the \ac{SNR} can reach above 20 and both the luminosity distance and the chirp mass can be determined \ccbx{at} the 10\% level.
This will enable the \ccbx{discrimination} of different seed models.
For relatively low redshift (e.g., $z=2$) sources, for which there is \ccbx{the chance of finding an} electromagnetic counterpart, TianQin has enough sensitivity to detect sources with chirp masses in the range $10^4\sim10^6~\Msun$ 24 hours before the final merger. Such \ccbx{detections} can have \acp{SNR} as large as 23 and the sky localization error less than 100 deg$^2$, and so can be used to trigger and guide the observation of electromagnetic instruments. More detail on the detection of \ac{MBHBs} with TianQin can be found in \cite{Wang:2019ryf,Feng:2019wgq}.

It is interesting to consider the possible scenario that TianQin is observing at the same time \ccbx{as} another detector, such as LISA. \ccbx{In this case the} simultaneous detection of a \ac{MBHB} merger signal can significantly improve the precision of source parameter estimation, such as the three-dimensional localisation as well as the merger time \cite{Feng:2020}.

\subsection{\ccbx{Extreme mass ratio inspirals (EMRIs)}}

Massive black holes in the center of galaxies \ccbx{in the local universe} can be accompanied by nuclear stellar clusters with sizes of a few parsec and masses up to $10^7\sim10^8~\Msun$ \cite{Schodel:2014wma}. Compact objects such as stellar mass black holes can sink into the gravitational potential well through two-body relaxation.
If \ccbx{they later become gravitationally bound} with the massive black hole, \ccbx{these} compact objects may eventually merge into the massive black hole through gravitational radiation.
The \ac{GW} signals from such events, called \ccbx{\acp{EMRI}}, have distinct features: they can last for hundreds of thousands of cycles and they stop abruptly if the central objects are really the Kerr black holes as predicted by \ac{GR}. The detection of such \ac{GW} signals with TianQin can precisely map the surrounding geometry of the central black holes and help test the Kerr hypothesis and \ac{GR} to exquisite precision \cite{Barack:2006pq}.
\ccrx{It has \ccbx{also} been shown that the detection of \acp{EMRI} with TianQin can be used to study boson stars \cite{Guo:2019sns}, to reveal new formation channels of \acp{EMRI} through the detection of their electromagnetic counterpart \cite{Wang:2019bbk}, and to further reveal the dynamics around the cental massive black hole in our own Galaxy \cite{Han:2020dql}.}

The rate of \acp{EMRI} depends on a variety of astrophysical processes determining the evolution of \ccbx{massive black holes} and the surrounding compact objects. A dozen population models have been developed by Babak et al. \cite{Babak:2017tow}. The calculation of \ccbx{accurate \ac{EMRI} waveforms} is the most challenging among all \ac{GW} sources. Despite a lot of progress, the problem to accurately and efficiently generate full \ac{EMRI} waveforms with the effect of self-force \ccbx{is still open} \cite{Drasco:2005kz, vandeMeent:2016pee,Warburton:2014bya,Burko:2006ua, Nagar:2006xv,Huerta:2008gb,Han:2016zee}.
\ccrx{In this regard, kludge \ac{EMRI} waveforms have been used to reduced the computational burden \cite{Fan:2020zhy,Babak:2017tow}.
With additional uncertainties in the plunge time of \ccbx{\ac{EMRI} waveforms}, it has been found that dozens to thousands of \acp{EMRI} may be detected with TianQin, with the horizon distance marked by sources with masses \ccbx{of} order  $10^6~\Msun$ at redshifts near $z=2$.}
For the parameters that strongly affect the phases of \acp{GW}, the \ccbx{relative} precision of parameter estimation is typically of the order $10^{-6}$. For other parameters, such as the luminosity distance and the sky localization, the (relative) uncertainty can be constrained to better than $\Delta D_L/D_L\sim 10\%$ and to the level $\Delta\Omega_S \sim10^{-3}$ deg$^2$, respectively, for the majority of sources.
The determination of the three dimensional location is precise enough so \ccbx{that} the detected \acp{EMRI} can be used as standard sirens for \ccbx{cosmological} study \cite{Schutz:1986gp,Tamanini:2016zlh}.
More detail on the detection of \acp{EMRI} with TianQin can be found in \cite{Fan:2020zhy}.

\subsection{Various cosmological processes}

In addition to the late-stage astronomical objects \ccrx{described in the preceding subsections}, many processes related to the (very) early universe can produce a \ac{SGWB}.\footnote{\ccrx{\ac{SGWB} can also have other origins, such as purely astronomical processes, and even interactions of massive black holes with ultralight bosons \cite{Brito:2017wnc,Brito:2017zvb}.}}
The main generating mechanisms include the amplification of vacuum \ccbx{fluctuations} re-entering the Hubble horizon during inflation, post-inflationary preheating and related non-perturbative phenomena, \ccbx{first-order phase transitions} (PTs) of the early universe and cosmic defects due to \ccbx{symmetry-breaking} of topological structure \ccrx{(see \cite{Caprini:2018mtu} for a recent review)}.
\ccrx{Recent studies have analyzed the important consequences of the first-order PTs that are triggered by introducing an effective operator~\cite{Wang:2020jrd}, by the extended Higgs sector~\cite{Alves:2018oct,Chen:2019ebq,Zhou:2020ojf,Alves:2018jsw,Zhou:2020xqi,Wang:2019pet} as well as by other exotic TeV scale particles (i.e., axion-like particles~\cite{Dev:2019njv,Zhou:2020stj}, heavy neutrinos ~\cite{Bian:2019szo,DiBari:2020bvn} and composite resonances~\cite{Bian:2019kmg}) and subsequently \ccbx{have} assessed the potential of using TianQin to detect the SGWB produced from these processes.
In some well-motivated scenarios, TianQin can detect \ccbx{the} \ac{SGWB} of cosmological origin with an optimal \ac{SNR} as high as the order of $10^5$~\cite{Liang:2020sto}. So TianQin has the potential to probe energy scales above the reach of near-future particle accelerators~\cite{Dev:2019njv,Bian:2019szo,DiBari:2020bvn,Bian:2019kmg} and even up to the Grand Unified Theories scale~\cite{Huang:2020bbe}.}

\ccrx{The formation of cosmic string \ccbx{networks} during the process of symmetry-breaking also produces \ccbx{a} SGWB, which can be an important observational effect of Grand Unified Theories~\cite{King:2020hyd}.
In some idealistic cases, TianQin may be able to probe the general network of cosmic defects with tensions $G\mu \gtrsim \mathcal{O}(10^{-6})$ and particularly the cosmic strings with $G\mu \gtrsim \mathcal{O}(10^{-17})$, corresponding to the scale of symmetry-breaking at $10^{16}{\rm ~GeV}$ and $10^{10}{\rm ~GeV}$, respectively~\cite{Liang:2020sto}.
Finally, the detectability of secondary GW backgrounds produced during inflation and from the PBH after inflation has been evaluated in \cite{Gong:2017qlj,Lu:2019sti,Lin:2020goi,Drees:2019xpp}.}

Therefore, if any of these GW backgrounds is detected in \ccbx{the} future, it will provide crucial information on the cosmic history and place constraints on the fundamental theories describing the early universe or on the low-energy effective theories related to particle physics around the TeV scale.
\ccrx{More details of using TianQin to detect \ac{SGWB} will be given in \cite{Liang:2020sto}, which also contains \ccbx{a} detailed explanation of how to use the TianQin sensitivity curve in this regard.}

\subsection{Cosmology and fundamental physics}

\ac{GW} signals detected with TianQin can be used to fill in gaps in the expansion history of the universe. Important information can come from detecting \ac{SBHBs} within a redshift \ccbx{of order} $0.1$, \acp{EMRI} within a redshift \ccbx{of order} 1 and the merger of \ac{MBHBs} at other high redshifts. The detection/null detection of primordial \acp{GW} can offer valuable information on the history of the early universe \ccrx{\cite{Gong:2017qlj,Lu:2019sti,Lin:2020goi}}. It has been found that under ideal circumstances, the observation of massive black hole binary mergers can pinpoint the Hubble constant to a precision of the order of $10^{-2}$ \cite{Wang:2019tto,Zhu:2020}.

Every aspect of \acp{GW} can be used to test \ac{GR}.
At the stage of generation, \acp{GW} from \ac{GCBs} are best used for testing the extra radiation channels and extra polarization modes of \acp{GW}\footnote{\ccrx{A devoted study on the extra polarizations of \acp{GW} from different gravitational theories and the corresponding responses and sensitivities of TianQin can be found in \cite{Hou:2017bqj,Gong:2018cgj,Gong:2018ybk,Gong:2018vbo,Liang:2019pry,Zhang:2020khm,Zhang:2019oet}.}}\ccbx{;}
\ac{SBHBs} are ideal sources for multi-band observations that can help \ccbx{improve by orders of magnitudes the constraints on certain parameters describing deviations from \ac{GR} \cite{Barausse:2016eii,Sesana:2016ljz};}
the \ccbx{mergers} of \ac{MBHBs} are ideal sources for testing various post-Einstein parameters and the Kerr nature of black holes\ccbx{;}
and \acp{EMRI} can be used to study the surrounding geometry of a massive black hole to high precision.
At the propagation stage, \ac{GW} signals from far away sources can be used to \ccbx{constrain} the \ccbx{dispersion} relation and the speed of \ac{GW} propagation.
\ccbx{As a quantitative analysis of how TianQin can perform,} a test of the black hole no-hair theorem and a study of the constraint on a particular modified gravity theory using \ccbx{ringdown signals} from the merger of \ac{MBHBs} have been carried out in \cite{Shi:2019hqa} and \cite{Bao:2019kgt}, respectively.

\section{Roadmap and technology progress}

\ccrx{Inertial reference and inter-satellite laser interferometry are two core technologies of TianQin, and the corresponding technology \ccbx{requirements}, grossly characterised by $S_a$ and $S_x$ in Table \ref{tab:configuration}, are} \ccbx{prerequisites} to achieving the scientific goals discussed in the last section.

As an important foundation to meet such requirements, the study and development of inertial sensors based on capacitive sensing and electrostatic control have \ccbx{been underway} since 2000 at the Centre for Gravitational Experiments, Huazhong University of Science and Technology (CGE-HUST).
Several \ccbx{flight} models have been constructed and successively tested in orbit \cite{s17091943}.
Efforts are being made to further improve on the performance of the instruments \ccrx{\cite{Pei_2019,Yang:2020dgd}.}
A study of the effect of space plasma on the motion of test masses has been carried out in \cite{Su:2020ifb}.
The development of high precision laser interferometers has \ccbx{been underway} since 2002.
A demonstration system of laser interferometer with 10-m armlength has been built in 2011 \cite{Yeh:2011rsi} and a prototype of transponder-type inter-satellite laser interferometer has been constructed and tested in 2015 \cite{Liang:2015rsi,luo:2015rsi}.
\ccrx{Further efforts in this direction include the study of the inter-satellite laser link acquisition \cite{Yeh:2018rsi} and \ccbx{sampling} jitter noise of the digital phasemeter \cite{Liang:2018rsi}.
Relativistic \ccbx{effects} on the laser propagation and clock comparison have been studied in \cite{Qin:2019ams,Qin:2019qqx}, while a large-scale passive laser gyroscope aiming to help link the celestial and terrestrial reference frames is being developed \cite{Zhang:2020rsi,Zhang:2020gux}.}

In order to systematically bring the key technologies of TianQin to maturity, a technology roadmap called the 0123 plan has been adopted since the beginning of the TianQin project:
\begin{itemize}
\item Step 0: Acquiring the capability to obtain high precision orbit information for satellites in the TianQin orbit through lunar laser ranging experiments.
\item Step 1: \ccbx{Using single satellite missions, where the} main goal is to test and demonstrate the maturity of the inertial reference technology;
\item Step 2: \ccbx{Using a mission with a pair of satellites, where the} main goal is to test and demonstrate the maturity of the inter-satellite laser interferometry technology;
\item Step 3: Launching a \ccbx{constellation} of three satellites to form the space-based GW observatory, TianQin.
\end{itemize}
At each step, one to several independent missions/projects are expected, depending on the progress of  technologies and the opportunities of space missions, and the numbers labelling the steps are the numbers of dedicated satellites \ccbx{which} need to be built for each independent mission/project of the step.

In the following, we report on two major projects, one in step 0 and the other in step 1, that are being carried out.

\subsection{Lunar Laser Ranging (LLR)}

In order to help better determine the positions of the TianQin satellites, Lunar Laser Ranging (LLR) has been an important part of the TianQin project since early on. Because no dedicated satellite needs to be built for this part of the work, all the projects related to LLR are categorized as step 0.

A major project on LLR was jointly approved and supported by the China National Space Administration (CNSA) and the National Natural Science Foundation of China (NSFC) in 2016. The project mainly involves upgrading/constructing laser ranging stations on the ground and creating a new generation of corner-cube retro-reflectors (CCRs) to be installed on the lunar relay satellite, QueQiao, for the Chang'E 4 mission.

Through the project, the LLR station of the Yunnan Observatories in \ccbx{Kunming} has been upgraded and on 22 January 2018 the station became the first in China to have successfully ranged to the \ccbx{Moon} \cite{LiYeh:2019}.
A new laser ranging station equipped with a $1.2~\um$ telescope has also been constructed on \ccbx{Fenghuang mountain near} the Zhuhai Campus of Sun Yat-sen University. The station has successfully received laser ranging signals from all \ccbx{five retro-reflectors} on the Moon. A single-body hollow CCR with $17~\ucm$ aperture has also been created and \ccbx{was} launched with the QueQiao satellite on 21 May 2018 \cite{HeLiu:2018}.

\subsection{The TianQin-1 Satellite (TQ-1)}

The major objectives of TQ-1 include testing the technologies of inertial sensing, micro-Newton propulsion, drag free control, laser interferometry, temperature control and center-of-mass measurement with in-orbit experiments.

The preparation for TQ-1 started in 2016 and the project received official approval from CNSA in 2018. \ccbx{Support} for TQ-1 \ccbx{has} also been provided by the Ministry of Education, the Guangdong Provincial Government and the Zhuhai Municipal Government of the People's Republic of China. TQ-1 was successfully launched on 20 December 2019 from the Taiyuan Satellite Launch Center in north China's Shanxi Province. The satellite completed its startup phase on 21 December 2019 and has been functioning smoothly since \ccbx{then}.

\ccbx{Results show} that the satellite has exceeded all of its mission requirements: by using the inertial sensor, which has a sensitivity \ccbx{of} $5\times10^{-12}~\um~\us^{-2}/\uHz^{1/2}$ at $0.1~\uHz\,$, as the key tool, the acceleration of the satellite has been measured and found to be about $1\times10^{-10}~\um ~\us^{-2}/\uHz^{1/2}$ at $0.1~\uHz\,$ and about $5\times10^{-11}~\um ~\us^{-2}/\uHz^{1/2}$ at $0.05~\uHz\,$.
\ccbx{The} performance of the micro-Newton thrusters has been evaluated and the thrust resolution is found to be about $0.1~\muN$ while the thrust noise is found to be about $0.3~\muN/\uHz^{1/2}$ at $0.1~\uHz$. \ccbx{The} residual noise of the satellite after drag free control is measured and found to be about $3\times10^{-9} ~\um~\us^{-2}/\uHz^{1/2}$ at $0.1~\uHz\,$, \ccrx{which mainly comes from the micro-Newton thrusters.} \ccbx{The} mismatch between the center-of-mass of the satellite and that of the test mass has also been measured with a precision better than $0.1~\umm$; the noise level of the optical readout system is about $30~\upm/\uHz^{1/2}$ at $0.1~\uHz\,$; the temperature stability at key temperature monitoring positions has been controlled to about $\pm3~\umK$ per orbit (about 97.13 min). More details on the performance of TQ-1 can be found in \cite{Luo:2020bls}.

By providing \ccbx{first hand data} on the in-orbit performance of the key payloads that are essential to the TianQin project, and by narrowing/quantifying the gap between the current technology capability and the requirement of TianQin, TQ-1 has marked a new milestone in the development of the TianQin project.

\subsection{Other developments}

\subsubsection{Dedicated research facilities}

The TianQin project involves constructing a few dedicated research facilities:
\begin{itemize}
\item The TianQin research building: to be used for research and technology development for the TianQin project. The building has been finished with a total area of more than 37 thousand \ccbx{square meters} and is scheduled to open in 2020.
\item The TianQin cave lab: to be used for research and technology development for the TianQin project. \ccbx{Excavation of the cave lab tunnel has started in 2019 and the tunnel was holed through on 5 June 2020.}
\item The Ground Simulation Facility (GSF): to be used for the integrated test and research on \ccbx{TianQin technologies and prototypes}. A pre-study project has been approved for the construction of \ccbx{this} facility.
\end{itemize}

\subsubsection{International collaboration}

International collaboration is an important aspect of the TianQin project. There have been 6 International Workshops on the TianQin Science Mission since 2014. On 18 December 2018, \ccrx{the TianQin Collaboration, including its international advisory committee,} was formally established during the fifth TianQin workshop.

\section*{Acknowledgments}

This work was supported in part by the Guangdong Major Project of Basic and Applied Basic Research (Grant No. 2019B030302001) and the National Natural Science Foundation of China (Grants No. 11703098, 11805286, 11805287, 11975319, 11654004, 11655001, 41811530087, 11690022). David Blair and Gerhard Heinzel have provided helpful comments and suggestions when the paper is being prepared.

\section*{References}
\bibliographystyle{iopart-num}
\bibliography{TQ-prg}

\providecommand{\newblock}{}
\begin{thebibliography}{100}
\expandafter\ifx\csname url\endcsname\relax
  \def\url#1{{\tt #1}}\fi
\expandafter\ifx\csname urlprefix\endcsname\relax\def\urlprefix{URL }\fi
\providecommand{\eprint}[2][]{\url{#2}}

\bibitem{Audley:2017drz}
Amaro-Seoane P {\em et~al.\/} (LISA) 2017  (\textit{Preprint}
  \eprint{1702.00786})

\bibitem{Sato:2017dkf}
Sato S {\em et~al.\/} 2017 {\em J. Phys. Conf. Ser.\/} {\bf 840} 012010

\bibitem{Isoyama:2018rjb}
Isoyama S {\em et~al.\/} 2018 {\em PTEP\/} {\bf 2018} 073E01 (\textit{Preprint}
  \eprint{1802.06977})

\bibitem{Luo:2015ght}
Luo J {\em et~al.\/} (TianQin) 2016 {\em Class. Quant. Grav.\/} {\bf 33} 035010
  (\textit{Preprint} \eprint{1512.02076})

\bibitem{Hellings:1996vz}
Hellings R 1996 {\em Contemp. Phys.\/} {\bf 37} 457--469

\bibitem{2012cosp...39.1890S}
Stebbins R {\em et~al.\/} 2012 {NASA's Gravitational-Wave Mission Concept
  Study} {\em 39th COSPAR Scientific Assembly\/} vol~39 p 1890

\bibitem{Tinto:2014eua}
Tinto M, DeBra D, Buchman S and Tilley S 2015 {\em Rev. Sci. Instrum.\/} {\bf
  86} 014501 (\textit{Preprint} \eprint{1410.1813})

\bibitem{Ye:2019txh}
Ye B~B, Zhang X, Zhou M~Y, Wang Y, Yuan H~M, Gu D, Ding Y, Zhang J, Mei J and
  Luo J 2019 {\em Int. J. Mod. Phys. D\/} {\bf 28} 1950121

\bibitem{Hu:2018yqb}
Hu X~C, Li X~H, Wang Y, Feng W~F, Zhou M~Y, Hu Y~M, Hu S~C, Mei J~W and Shao
  C~G 2018 {\em Class. Quant. Grav.\/} {\bf 35} 095008 (\textit{Preprint}
  \eprint{1803.03368})

\bibitem{doi:10.1142/S021827182050056X}
Tan Z, Ye B and Zhang X 2020 {\em International Journal of Modern Physics D\/}
  2050056

\bibitem{Strohmayer:2005uc}
Strohmayer T~E 2005 {\em Astrophys. J.\/} {\bf 627} 920--925 (\textit{Preprint}
  \eprint{astro-ph/0504150})

\bibitem{Wang:2019ryf}
Wang H~T {\em et~al.\/} 2019 {\em Phys. Rev. D\/} {\bf 100} 043003
  (\textit{Preprint} \eprint{1902.04423})

\bibitem{Feng:2019wgq}
Feng W~F, Wang H~T, Hu X~C, Hu Y~M and Wang Y 2019 {\em Phys. Rev. D\/} {\bf
  99} 123002 (\textit{Preprint} \eprint{1901.02159})

\bibitem{Shi:2019hqa}
Shi C, Bao J, Wang H, Zhang J~d, Hu Y, Sesana A, Barausse E, Mei J and Luo J
  2019 {\em Phys. Rev. D\/} {\bf 100} 044036 (\textit{Preprint}
  \eprint{1902.08922})

\bibitem{Bao:2019kgt}
Bao J, Shi C, Wang H, Zhang J~d, Hu Y, Mei J and Luo J 2019 {\em Phys. Rev.
  D\/} {\bf 100} 084024 (\textit{Preprint} \eprint{1905.11674})

\bibitem{Liu:2020eko}
Liu S, Hu Y~M, Zhang J~d and Mei J 2020 {\em Phys. Rev. D\/} {\bf 101} 103027
  (\textit{Preprint} \eprint{2004.14242})

\bibitem{Huang:2020rjf}
Huang S~J, Hu Y~M, Korol V, Li P~C, Liang Z~C, Lu Y, Wang H~T, Yu S and Mei J
  2020  (\textit{Preprint} \eprint{2005.07889})

\bibitem{Fan:2020zhy}
Fan H~M, Hu Y~M, Barausse E, Sesana A, Zhang J~D, Zhang X, Zi T~G and Mei J
  2020  (\textit{Preprint} \eprint{2005.08212})

\bibitem{Liang:2020sto}
Liang Z~C {\em et~al.\/} {\em In preparation.\/}

\bibitem{Zhu:2020}
Zhu L {\em et~al.\/} {\em In preparation.\/}

\bibitem{Lu:2019log}
Lu X~Y, Tan Y~J and Shao C~G 2019 {\em Phys. Rev. D\/} {\bf 100} 044042

\bibitem{Hu:2017yoc}
Hu Y~M, Mei J and Luo J 2017 {\em Natl. Sci. Rev.\/} {\bf 4} 683--684

\bibitem{Evans:1987qa}
Evans C~R {\em et~al.\/} 1987 {\em Astrophys.\ J.\/} {\bf 323} 129--139

\bibitem{1987A&A...176L...1L}
Lipunov V~M {\em et~al.\/} 1987 {\em Astron. Astrophys.\/} {\bf 176} L1--L4

\bibitem{Hils:1990vc}
Hils D {\em et~al.\/} 1990 {\em Astrophys.\ J.\/} {\bf 360} 75--94

\bibitem{Nelemans:2013yg}
Nelemans G 2013 {\em ASP Conf. Ser.\/} {\bf 467} 27--36 (\textit{Preprint}
  \eprint{1302.0138})

\bibitem{Rebassa-Mansergas:2018mtp}
Rebassa-Mansergas A, Toonen S, Korol V and Torres S 2019 {\em Mon. Not. Roy.
  Astron. Soc.\/} {\bf 482} 3656--3668 (\textit{Preprint} \eprint{1809.07158})

\bibitem{Adams:2012qw}
Adams M~R, Cornish N~J and Littenberg T~B 2012 {\em Phys. Rev. D\/} {\bf 86}
  124032 (\textit{Preprint} \eprint{1209.6286})

\bibitem{Korol:2018wep}
Korol V, Rossi E~M and Barausse E 2019 {\em Mon. Not. Roy. Astron. Soc.\/} {\bf
  483} 5518--5533 (\textit{Preprint} \eprint{1806.03306})

\bibitem{Wilhelm:2020qjc}
Wilhelm M~J, Korol V, Rossi E~M and D'Onghia E 2020  (\textit{Preprint}
  \eprint{2003.11074})

\bibitem{Littenberg:2018xxx}
Littenberg T~B and Yunes N 2019 {\em Class. Quant. Grav.\/} {\bf 36} 095017
  (\textit{Preprint} \eprint{1811.01093})

\bibitem{Philippoz:2017ywb}
Philippoz L and Jetzer P 2017 {\em J. Phys. Conf. Ser.\/} {\bf 840} 012057

\bibitem{Wang:2020pzc}
Wang B and Liu D 2020  (\textit{Preprint} \eprint{2005.01880})

\bibitem{Rodriguez:2019ucq}
Rodriguez J, Rueda J, Ruffini R, Zuluaga J, Blanco-Iglesias J and Loren-Aguilar
  P 2019  (\textit{Preprint} \eprint{1907.10532})

\bibitem{Brown:2011gq}
Brown W~R, Kilic M, Hermes J, Allende~Prieto C, Kenyon S~J and Winget D 2011
  {\em Astrophys. J. Lett.\/} {\bf 737} L23 (\textit{Preprint}
  \eprint{1107.2389})

\bibitem{kil14}
{Kilic} M, {Brown} W~R, {Gianninas} A, {Hermes} J~J, {Allende Prieto} C and
  {Kenyon} S~J 2014 {\em \mnras\/} {\bf 444} L1--L5 (\textit{Preprint}
  \eprint{1406.3346})

\bibitem{kup18}
{Kupfer} T, {Korol} V, {Shah} S, {Nelemans} G, {Marsh} T~R, {Ramsay} G, {Groot}
  P~J, {Steeghs} D~T~H and {Rossi} E~M 2018 {\em \mnras\/} {\bf 480} 302--309
  (\textit{Preprint} \eprint{1805.00482})

\bibitem{ram18}
{Ramsay} G, {Green} M~J, {Marsh} T~R, {Kupfer} T, {Breedt} E, {Korol} V,
  {Groot} P~J, {Knigge} C, {Nelemans} G, {Steeghs} D, {Woudt} P and
  {Aungwerojwit} A 2018 {\em \aap\/} {\bf 620} A141 (\textit{Preprint}
  \eprint{1810.06548})

\bibitem{bur19}
{Burdge} K, {Coughlin} M~W, {Fuller} J, {Gaensicke} B~T, {Kaplan} D~L,
  {Kulkarni} S~R, {Marsh} T~R, {Prince} T~A and {Toloza Castillo} O~F 2019 {ZTF
  J1539+5027: the Shortest Period Eclipsing White Dwarf Binary} HST Proposal

\bibitem{LIGOScientific:2018mvr}
Abbott B {\em et~al.\/} (LIGO Scientific, Virgo) 2019 {\em Phys. Rev. X\/} {\bf
  9} 031040 (\textit{Preprint} \eprint{1811.12907})

\bibitem{Colpi:2016fup}
Colpi M and Sesana A 2017 {\em {Gravitational Wave Sources in the Era of
  Multi-Band Gravitational Wave Astronomy}\/} pp 43--140 (\textit{Preprint}
  \eprint{1610.05309})

\bibitem{Sesana:2016ljz}
Sesana A 2016 {\em Phys. Rev. Lett.\/} {\bf 116} 231102 (\textit{Preprint}
  \eprint{1602.06951})

\bibitem{Chen:2017gfm}
Chen X and Amaro-Seoane P 2017 {\em Astrophys. J. Lett.\/} {\bf 842} L2
  (\textit{Preprint} \eprint{1702.08479})

\bibitem{Samsing:2019dtb}
Samsing J, D'Orazio D~J, Kremer K, Rodriguez C~L and Askar A 2020 {\em Phys.
  Rev. D\/} {\bf 101} 123010 (\textit{Preprint} \eprint{1907.11231})

\bibitem{Randall:2019znp}
Randall L and Xianyu Z~Z 2019  (\textit{Preprint} \eprint{1907.02283})

\bibitem{Carson:2019yxq}
Carson Z and Yagi K 2019 {\em MDPI Proc.\/} {\bf 17} 5 (\textit{Preprint}
  \eprint{1908.07103})

\bibitem{LIGOScientific:2018jsj}
Abbott B {\em et~al.\/} (LIGO Scientific, Virgo) 2019 {\em Astrophys. J.\/}
  {\bf 882} L24 (\textit{Preprint} \eprint{1811.12940})

\bibitem{1971MNRAS.152..461L}
{Lynden-Bell} D and {Rees} M~J 1971 {\em Monthly Notices of the Royal
  Astronomical Society\/} {\bf 152} 461

\bibitem{1982MNRAS.200..115S}
{Soltan} A 1982 {\em Monthly Notices of the Royal Astronomical Society\/} {\bf
  200} 115--122

\bibitem{Kormendy:1995er}
Kormendy J and Richstone D 1995 {\em Ann. Rev. Astron. Astrophys.\/} {\bf 33}
  581

\bibitem{Gultekin:2009qn}
Gultekin K {\em et~al.\/} 2009 {\em Astrophys. J.\/} {\bf 698} 198--221
  (\textit{Preprint} \eprint{0903.4897})

\bibitem{Berti:2018cxi}
Berti E, Yagi K and Yunes N 2018 {\em Gen. Rel. Grav.\/} {\bf 50} 46
  (\textit{Preprint} \eprint{1801.03208})

\bibitem{Liu:2020ddo}
Liu H, Zhang C, Gong Y, Wang B and Wang A 2020  (\textit{Preprint}
  \eprint{2002.06360})

\bibitem{Stott:2018opm}
Stott M~J and Marsh D~J 2018 {\em Phys. Rev. D\/} {\bf 98} 083006
  (\textit{Preprint} \eprint{1805.02016})

\bibitem{Woods:2018lty}
Woods T~E {\em et~al.\/} 2019 {\em Publ. Astron. Soc. Austral.\/} {\bf 36} e027
  (\textit{Preprint} \eprint{1810.12310})

\bibitem{Barausse:2012fy}
Barausse E 2012 {\em Mon. Not. Roy. Astron. Soc.\/} {\bf 423} 2533--2557
  (\textit{Preprint} \eprint{1201.5888})

\bibitem{Klein:2015hvg}
Klein A {\em et~al.\/} 2016 {\em Phys. Rev. D\/} {\bf 93} 024003
  (\textit{Preprint} \eprint{1511.05581})

\bibitem{Barausse:2020mdt}
Barausse E, Dvorkin I, Tremmel M, Volonteri M and Bonetti M 2020
  (\textit{Preprint} \eprint{2006.03065})

\bibitem{Jiang_2019}
Jiang Z, Wang J, Gao L, Zhang F~H, Guo Q, Wang L and Pan J 2019 {\em Research
  in Astronomy and Astrophysics\/} {\bf 19} 151 ISSN 2397-6209

\bibitem{Feng:2020}
Feng W {\em et~al.\/} {\em To be submitted\/}

\bibitem{Schodel:2014wma}
{Sch\"odel, R and Feldmeier, A and Neumayer, N and Meyer, L and Yelda, S} 2014
  {\em Class. Quant. Grav.\/} {\bf 31} 244007 (\textit{Preprint}
  \eprint{1411.4504})

\bibitem{Barack:2006pq}
Barack L and Cutler C 2007 {\em Phys. Rev. D\/} {\bf 75} 042003
  (\textit{Preprint} \eprint{gr-qc/0612029})

\bibitem{Guo:2019sns}
Guo H~K, Sinha K and Sun C 2019 {\em JCAP\/} {\bf 09} 032 (\textit{Preprint}
  \eprint{1904.07871})

\bibitem{Wang:2019bbk}
Wang Y, Wang F, Zou Y and Dai Z 2019 {\em Astrophys. J. Lett.\/} {\bf 886} L22
  (\textit{Preprint} \eprint{1911.04117})

\bibitem{Han:2020dql}
Han W~B, Zhong X~Y, Chen X and Xin S 2020  (\textit{Preprint}
  \eprint{2004.04016})

\bibitem{Babak:2017tow}
Babak S, Gair J, Sesana A, Barausse E, Sopuerta C~F, Berry C~P, Berti E,
  Amaro-Seoane P, Petiteau A and Klein A 2017 {\em Phys. Rev. D\/} {\bf 95}
  103012 (\textit{Preprint} \eprint{1703.09722})

\bibitem{Drasco:2005kz}
Drasco S and Hughes S~A 2006 {\em Phys. Rev. D\/} {\bf 73} 024027 [Erratum:
  Phys.Rev.D 88, 109905 (2013), Erratum: Phys.Rev.D 90, 109905 (2014)]
  (\textit{Preprint} \eprint{gr-qc/0509101})

\bibitem{vandeMeent:2016pee}
van~de Meent M 2016 {\em Phys. Rev. D\/} {\bf 94} 044034 (\textit{Preprint}
  \eprint{1606.06297})

\bibitem{Warburton:2014bya}
Warburton N 2015 {\em Phys. Rev. D\/} {\bf 91} 024045 (\textit{Preprint}
  \eprint{1408.2885})

\bibitem{Burko:2006ua}
Burko L~M and Khanna G 2007 {\em EPL\/} {\bf 78} 60005 (\textit{Preprint}
  \eprint{gr-qc/0609002})

\bibitem{Nagar:2006xv}
Nagar A, Damour T and Tartaglia A 2007 {\em Class. Quant. Grav.\/} {\bf 24}
  S109--S124 (\textit{Preprint} \eprint{gr-qc/0612096})

\bibitem{Huerta:2008gb}
Huerta E and Gair J~R 2009 {\em Phys. Rev. D\/} {\bf 79} 084021 [Erratum:
  Phys.Rev.D 84, 049903 (2011)] (\textit{Preprint} \eprint{0812.4208})

\bibitem{Han:2016zee}
Han W~B 2016 {\em Class. Quant. Grav.\/} {\bf 33} 065009 (\textit{Preprint}
  \eprint{1609.06817})

\bibitem{Schutz:1986gp}
Schutz B~F 1986 {\em Nature\/} {\bf 323} 310--311

\bibitem{Tamanini:2016zlh}
Tamanini N, Caprini C, Barausse E, Sesana A, Klein A and Petiteau A 2016 {\em
  JCAP\/} {\bf 04} 002 (\textit{Preprint} \eprint{1601.07112})

\bibitem{Brito:2017wnc}
Brito R, Ghosh S, Barausse E, Berti E, Cardoso V, Dvorkin I, Klein A and Pani P
  2017 {\em Phys. Rev. Lett.\/} {\bf 119} 131101 (\textit{Preprint}
  \eprint{1706.05097})

\bibitem{Brito:2017zvb}
Brito R, Ghosh S, Barausse E, Berti E, Cardoso V, Dvorkin I, Klein A and Pani P
  2017 {\em Phys. Rev. D\/} {\bf 96} 064050 (\textit{Preprint}
  \eprint{1706.06311})

\bibitem{Caprini:2018mtu}
Caprini C and Figueroa D~G 2018 {\em Class. Quant. Grav.\/} {\bf 35} 163001
  (\textit{Preprint} \eprint{1801.04268})

\bibitem{Wang:2020jrd}
Wang X, Huang F~P and Zhang X 2020 {\em JCAP\/} {\bf 05} 045 (\textit{Preprint}
  \eprint{2003.08892})

\bibitem{Alves:2018oct}
Alves A, Ghosh T, Guo H~K and Sinha K 2018 {\em JHEP\/} {\bf 12} 070
  (\textit{Preprint} \eprint{1808.08974})

\bibitem{Chen:2019ebq}
Chen N, Li T, Wu Y and Bian L 2020 {\em Phys. Rev. D\/} {\bf 101} 075047
  (\textit{Preprint} \eprint{1911.05579})

\bibitem{Zhou:2020ojf}
Zhou R, Yang J and Bian L 2020 {\em JHEP\/} {\bf 04} 071 (\textit{Preprint}
  \eprint{2001.04741})

\bibitem{Alves:2018jsw}
Alves A, Ghosh T, Guo H~K, Sinha K and Vagie D 2019 {\em JHEP\/} {\bf 04} 052
  (\textit{Preprint} \eprint{1812.09333})

\bibitem{Zhou:2020xqi}
Zhou R and Bian L 2020  (\textit{Preprint} \eprint{2001.01237})

\bibitem{Wang:2019pet}
Wang X, Huang F~P and Zhang X 2020 {\em Phys. Rev. D\/} {\bf 101} 015015
  (\textit{Preprint} \eprint{1909.02978})

\bibitem{Dev:2019njv}
Dev P~B, Ferrer F, Zhang Y and Zhang Y 2019 {\em JCAP\/} {\bf 11} 006
  (\textit{Preprint} \eprint{1905.00891})

\bibitem{Zhou:2020stj}
Zhou Z, Yan J, Addazi A, Cai Y~F, Marciano A and Pasechnik R 2020
  (\textit{Preprint} \eprint{2003.13244})

\bibitem{Bian:2019szo}
Bian L, Cheng W, Guo H~K and Zhang Y 2019  (\textit{Preprint}
  \eprint{1907.13589})

\bibitem{DiBari:2020bvn}
Di~Bari P, Marfatia D and Zhou Y~L 2020  (\textit{Preprint}
  \eprint{2001.07637})

\bibitem{Bian:2019kmg}
Bian L, Wu Y and Xie K~P 2019 {\em JHEP\/} {\bf 12} 028 (\textit{Preprint}
  \eprint{1909.02014})

\bibitem{Huang:2020bbe}
Huang W~C, Sannino F and Wang Z~W 2020  (\textit{Preprint} \eprint{2004.02332})

\bibitem{King:2020hyd}
King S~F, Pascoli S, Turner J and Zhou Y~L 2020  (\textit{Preprint}
  \eprint{2005.13549})

\bibitem{Gong:2017qlj}
Di H and Gong Y 2018 {\em JCAP\/} {\bf 07} 007 (\textit{Preprint}
  \eprint{1707.09578})

\bibitem{Lu:2019sti}
Lu Y, Gong Y, Yi Z and Zhang F 2019 {\em JCAP\/} {\bf 12} 031
  (\textit{Preprint} \eprint{1907.11896})

\bibitem{Lin:2020goi}
Lin J, Gao Q, Gong Y, Lu Y, Zhang C and Zhang F 2020 {\em Phys. Rev. D\/} {\bf
  101} 103515 (\textit{Preprint} \eprint{2001.05909})

\bibitem{Drees:2019xpp}
Drees M and Xu Y 2019  (\textit{Preprint} \eprint{1905.13581})

\bibitem{Wang:2019tto}
Wang L~F, Zhao Z~W, Zhang J~F and Zhang X 2019  (\textit{Preprint}
  \eprint{1907.01838})

\bibitem{Hou:2017bqj}
Hou S, Gong Y and Liu Y 2018 {\em Eur. Phys. J. C\/} {\bf 78} 378
  (\textit{Preprint} \eprint{1704.01899})

\bibitem{Gong:2018cgj}
Gong Y, Hou S, Liang D and Papantonopoulos E 2018 {\em Phys. Rev. D\/} {\bf 97}
  084040 (\textit{Preprint} \eprint{1801.03382})

\bibitem{Gong:2018ybk}
Gong Y and Hou S 2018 {\em Universe\/} {\bf 4} 85 (\textit{Preprint}
  \eprint{1806.04027})

\bibitem{Gong:2018vbo}
Gong Y, Hou S, Papantonopoulos E and Tzortzis D 2018 {\em Phys. Rev. D\/} {\bf
  98} 104017 (\textit{Preprint} \eprint{1808.00632})

\bibitem{Liang:2019pry}
Liang D, Gong Y, Weinstein A~J, Zhang C and Zhang C 2019 {\em Phys. Rev. D\/}
  {\bf 99} 104027 (\textit{Preprint} \eprint{1901.09624})

\bibitem{Zhang:2020khm}
Zhang C, Gao Q, Gong Y, Wang B, Weinstein A~J and Zhang C 2020 {\em Phys. Rev.
  D\/} {\bf 101} 124027 (\textit{Preprint} \eprint{2003.01441})

\bibitem{Zhang:2019oet}
Zhang C, Gao Q, Gong Y, Liang D, Weinstein A~J and Zhang C 2019 {\em Phys. Rev.
  D\/} {\bf 100} 064033 (\textit{Preprint} \eprint{1906.10901})

\bibitem{Barausse:2016eii}
Barausse E, Yunes N and Chamberlain K 2016 {\em Phys. Rev. Lett.\/} {\bf 116}
  241104 (\textit{Preprint} \eprint{1603.04075})

\bibitem{s17091943}
Bai Y~Z {\em et~al.\/} 2017 {\em Sensors\/} {\bf 17} ISSN 1424-8220

\bibitem{Pei_2019}
Pei S~X, Liu L, Wu S~C, Bai Y~Z and Zhou Z~B 2019 {\em Classical and Quantum
  Gravity\/} {\bf 36} 235023

\bibitem{Yang:2020dgd}
Yang F, Bai Y, Hong W, Li H, Liu L, Sumner T~J, Yang Q, Zhao Y and Zhou Z 2020
  {\em Class. Quant. Grav.\/} {\bf 37} 115005

\bibitem{Su:2020ifb}
Su W {\em et~al.\/} 2020  (\textit{Preprint} \eprint{2004.00254})

\bibitem{Yeh:2011rsi}
Yeh H~C, Yan Q~Z, Liang Y, Wang Y and Luo J 2011 {\em Review of Scientific
  Instruments\/} {\bf 82} 044501 -- 044501

\bibitem{Liang:2015rsi}
Liang Y, Duan H~Z, Xiao X~L, Wei B~B and Yeh H~C 2015 {\em The Review of
  scientific instruments\/} {\bf 86} 016106

\bibitem{luo:2015rsi}
Luo Y, Li H, Yeh H~C and Luo J 2015 {\em Review of Scientific Instruments\/}
  {\bf 86} 044501

\bibitem{Yeh:2018rsi}
Zhang J~Y, Ming M, Jiang Y~Z, Duan H~Z and Yeh H~C 2018 {\em Rev. Sci.
  Instrum.\/} {\bf 89} 064501

\bibitem{Liang:2018rsi}
Liang Y~R 2018 {\em Rev. Sci. Instrum.\/} {\bf 89} 036106

\bibitem{Qin:2019ams}
Qin C~G, Tan Y~J, Chen Y~F and Shao C~G 2019 {\em Phys. Rev. D\/} {\bf 100}
  064063

\bibitem{Qin:2019qqx}
Qin C~G, Tan Y~J and Shao C~G 2019 {\em Class. Quant. Grav.\/} {\bf 36} 055008

\bibitem{Zhang:2020rsi}
Zhang F, Liu K, Li Z, Cheng F, Feng X, Li K, Lu Z and Zhang J 2020 {\em Rev.
  Sci. Instrum.\/} {\bf 91} 013001

\bibitem{Zhang:2020gux}
Zhang F {\em et~al.\/} 2020  (\textit{Preprint} \eprint{2003.05310})

\bibitem{LiYeh:2019}
Li Y {\em et~al.\/} 2019 {\em Chinese J. of Lasers\/} {\bf 46} 0104004--1

\bibitem{HeLiu:2018}
He Y {\em et~al.\/} 2018 {\em Res.\ Astron.\ Astrop.\/} {\bf 18} 131--144

\bibitem{Luo:2020bls}
Luo J {\em et~al.\/} 2020 {\em Class. Quant. Grav.\/} {\bf 37} 185013
  (\textit{Preprint} \eprint{2008.09534})

\end{thebibliography}
\end{document}